\documentclass[]{aastex631}
\usepackage{csquotes,hyperref,pgffor}

\received{9 July 2021}
\revised{24 August 2021}
\accepted{17 September 2021}
\submitjournal{Acta Astronautica}

\shorttitle{SETI in 2020}
\shortauthors{Wright}
\graphicspath{{./}{figures/}}
\newcommand{\PSUAA}{Department of Astronomy \& Astrophysics, 525 Davey Laboratory, The Pennsylvania State University, University Park, PA, 16802, USA}
\newcommand{\PSUCEHW}{Center for Exoplanets and Habitable Worlds, 525 Davey Laboratory, The Pennsylvania State University, University Park, PA, 16802, USA}
\newcommand{\PSETI}{Penn State Extraterrestrial Intelligence Center, 525 Davey Laboratory, The Pennsylvania State University, University Park, PA, 16802, USA}

\newcommand{\mycounter}[2]{
\newcounter{#1}
\setcounter{#1}{#2}
}

\mycounter{Npapers}{0}
\mycounter{searches}{7}
\addtocounter{Npapers}{\thesearches}
\mycounter{methods}{16}
\addtocounter{Npapers}{\themethods}
\mycounter{targets}{8}
\addtocounter{Npapers}{\thetargets}
\mycounter{development}{15}
\addtocounter{Npapers}{\thedevelopment}
\mycounter{theory}{19}
\addtocounter{Npapers}{\thetheory}
\mycounter{social}{10}
\addtocounter{Npapers}{\thesocial}

\begin{document}

\title{SETI in 2020}

\correspondingauthor{Jason Wright}
\email{astrowright@gmail.com}

\author[0000-0001-6160-5888]{Jason T.\ Wright}
\affil{\PSUAA}
\affil{\PSUCEHW}
\affil{\PSETI}

\begin{abstract}

In the spirit of Trimble's ``Astrophysics in XXXX'' series, I very briefly and subjectively review developments in SETI in 2020. My primary focus is \theNpapers\ papers and books published or made public in 2020, which I sort into six broad categories: results from actual searches, new search methods and instrumentation, target and frequency selection, the development of technosignatures, theory of ETIs, and social aspects of SETI. 

\end{abstract}

\keywords{Search for Extraterrestrial Intelligence}

\section{Introduction} \label{sec:intro}
One of the very first citations I received to a first-author paper was by \citet{Trimble2005} in ``Astrophysics in 2004,'' part of the annual ``ApXX'' series which Virginia Trimble maintained from 1991--2006.  I always enjoyed reading her highly subjective take on what caught her eye in the  astrophysics literature the previous year.

SETI is a young field, and it strikes me that a similar effort would be able to span most of it in fewer pages than Trimble's reviews. Such a review could be closer to comprehensive, be a useful guide to the field on the latest developments, and also help give a sense for where the collective efforts of practitioners were being spent.

For this first installment of such a series, I have used the ADS SETI bibliography \citep{Reyes2019,LaFond2021} to find all of the works with publication dates in 2020. For the most part this means refereed papers published in their final form in 2020, but I also included some other entries where it seemed appropriate. I avoided compilations of previously published work, and saved some preprints appearing at the end of the year for the next installment. I ended up with \theNpapers\ papers to cite, though I may have missed some worthy work.

I have roughly categorized these papers into six categories: results from actual searches, new search methods and instrumentation, target and frequency seleciton, the development of technosignatures, theory of ETIs, and social aspects of SETI (including matters concerning METI). In what follows, I will follow Trimble's lead, and not strive for strict impartiality.

\section{SETI in 2020}

2020 was the first year of the pandemic, the lost year, the year of COVID. Even as I write this in 2021, many facilities are closed, and much of the world does not have access to the vaccines that have allowed many nations to begin re-opening and returning to normal ways of research. It is likely this review would have more papers to cite without this global tragedy.

Many opportunities for SETI practitioners to meet and interact were lost. For instance, the First Penn State SETI symposium, intended to be a major, annual, international meeting in the field, was postponed to 2021 (and then later to 2022). That symposium was also intended to inaugurate the Penn State Extraterrestrial Intelligence Center, dedicated to research, training, curriculum, and fundraising for the field.

But, being forced to interact remotely also made the world a little smaller.  The Technoclimes conference in August 2020 explored non-radio technosignatures and produced many new collaborations. Two talk series on technosignatures, the NASA Goddard Technosignatures Seminar (led by Ravi Kopparapu) and the PSETI Seminar, joined the IAA community meeting to create a trio of worldwide ``SETI colloquia'' that anyone could attend.

There was also movement on the funding front. The US Congress continued to indicate that technosignatures were properly part of NASA's portfolio (although no explicit funding was allocated), the Technoclimes workshop was supported by a NASA Exobiology grant, and two NASA Exoplanet Research Program grants were awarded to researchers studying technosignatures (Jean-Luc Margot at UCLA and Ann Marie Cody at the SETI Institute). Finally, the Characterizing Atmospheric Technosignatures (CATS) effort led by Adam Frank at the University of Rochester was announced, and its members joined the NExSS research coordination network at NASA.

2020 is also a notable landmark for SETI because it marks the target for which the ``SETI 2020'' roadmap was written \citep{SETI2020}.  We have probably made less progress than that work aspired to since then, but that work nonetheless proved influential in directing radio SETI efforts, for instance in the development of the Allen Telescope Array, but also for other efforts such as the Square Kilometer Array. 

2020 had more than its share of SETI-specific tragedy: in February we lost Freeman Dyson, one of the pioneers of the field, at the age of 96; and in December the instrument platform suspended above the 300-meter dish of the Arecibo radio telescope collapsed, destroying not only one of the most powerful and storied radio telescopes in the world, but an icon of the field.

\section{Results from Searches (\thesearches\ papers)}

The heart of SETI is the actual searches, and 2020 brought several new important results and upper limits.

The recent resurgence of the field is due in no small part to the Breakthrough Listen Initiative, which has injected funding, training, community, and energy into SETI.  The team and its collaborators published four new upper limits in the radio, making use of the extremely high bandwidth backends at Green Bank and Parkes. The first was by \citet{Price20}, reporting null detections for over 1300 stars over the entire sky in L and S bands (1--3.5 GHz). Following the methodology suggested by \citet{Heller16}, \citet{SheikhETZ} reported null results in C band (4--8 GHz) for stars in the ``restricted Earth Transit Zone,'' i.e.\ the region of the sky that ``sees'' Earth transit the Sun.  \citet{Wlodarczyk-Sroka2020} extended the upper limit calculations of previous searches to include \textit{all} of the sources in the beam (not just the target star at the center). And finally, \citet{Perez2020} reported null results for the Kepler-160 system.

Overall, this represents both a substantial increase in the ``Cosmic Haystack'' fraction searched, and the deployment of important new search strategies in the radio.

The Allen Telescope Array (ATA) continued its survey of the skies, publishing in 2020 a deep survey and null result of the Wow!\ Signal field. This work, by \citet{Harp2020a}, scanned the entire field consistent with the original signal, and included a 10 MHz bandwidth and $\sim$100 hours of observation, the longest follow-up yet performed. It is good to see the field followed up in this manner to search for potential repetitions.

\citet{Tremblay2020} reported on the deepest narrowband SETI search yet with the Murchison Widefield Array (MWA), this time of the Vela region.  These low-frequency searches explore very large regions of parameter space in the \citet{Haystack} Cosmic Haystack formalism, thanks to their sensitivity and very large fields of view. 

Perhaps the most offbeat and unconventional search of 2020 was that of \citet{Hippke2020a}, who followed a suggestion by \citet{Hsu2006} that there could be a message embedded in the cosmic microwave background, presumably put there intentionally at the observable universe's beginning. Not finding anything meaningful in the power spectrum, Hippke helpfully provides a 1000-bit representation of the CMB temperature spectrum for others to scrutinize for any messages that were not apparent to him.

\section{Search Methods and Instrumentation (\themethods\ papers)}

\subsection{New Instrumentation}

One of the most exciting projects on the horizon is PanoSETI, an all-sky optical/NIR fast transient project designed to search for pulsed laser emission.  2020 saw significant progress on the PanoSETI prototype at Palomar Observatory, as described in a trio of SPIE papers by \citet{Brown2020} (describing the overall mechanical system design) \citet{Liu2020} (describing the focal plane electronics and protocols), and \citet{Maire2020} (showing on-sky results and instrument design). 

In addition to being a SETI program, this project promises to open up new avenues for research in cosmic rays, as well as the potential for serendipitous discoveries in  almost completely unexplored regions of parameter space for fast optical and NIR transients.

Another exciting new SETI instrument is the Five-hundred-meter Aperture Spherical radio Telescope (FAST) and its SERENDIP VI multibeam spectrometer, which was declared fully operational in January.  \citet{Li2020b} described FAST's capabilities and its partnership with Breakthrough Listen, and \citet{Zhang2020} described the first observations and RFI mitigation strategies they employ. FAST is shaping up to be a worthy successor to Arecibo.

\subsection{Anomaly detection}

Following up on suggestions by \citet{Albrecht1988}, \citet{Djorgovski00}, and others to search for highly unusual events in large sky surveys, many SETI practitioners have begun work on anomaly-detection or outlier analysis as a way to identify unexpected, potentially technological phenomena.  \citet{Singam2020} provided an overview of searches for such ``non-canonical astrophysical phenomena'' and made suggestions for them.

\citet{Villarroel2020b} described the Vanishing and Appearing Sources during a Century of Observations project (VASCO), which leverages photographic and digital sky surveys across many decades to search for star-like sources that seem to disappear or appear.  The survey has many components, including hypothesis-driven searches (for instance for failed supernovae) and serendipitous searches driven by machine learning.  \citet{Villarroel2020a} describe the citizen science component of the project.

\citet{DelaTorre2020} explored the challenges of serendipitous detection with neural networks, and the role of human cognition and bias in such searches. He compares the results of a neural network to those of human volunteers identifying features in imagery, and uses the bright spots on Ceres as an example.

\citet{Brzycki2020} described using convolutional neural networks to identify and characterize narrow band signals in noisy data with an eye towards radio SETI. 
\citet{Lesnikowski2020} described an unsupervised neural network to analyze imagery of the moon to look for artifacts and other anomalies, and shows they can recover the obviously artificial Apollo 15 landing site with it.

Finally, \citet{Giles2020} discussed an outlier scoring technique for lightcurve data as a way to search for anomalies and new phenomena, and produce such scores for millions of light curves from the \textit{Kepler} mission. Such work is useful for finding anomalies such as Boyajian's star \citep{WTF}, which have proven interesting both for SETI and natural astrophysics purposes \citep{Wright16}.

\subsection{Other methodology}

\citet{Michaud2020} reviewed the case for preserving the lunar farside for radio astronomy and radio SETI, to take advantage of what is (today) a region free from terrestrial radio frequency interference. 

\citet{Sandora2020} outlined a Bayesian framework for designing and interpreting results from searches for biosignatures and technosignatures. \citet{Haqq-Misra2020a} argued that the relative abundance of technosignatures versus biosignatures detected in future astronomical surveys with space telescopes will help determine whether the ``Great Filter'' (which they define as the ``hardest step in planetary evolution'') is in our past or future. 

Finally, \citet{Seto2020} presented an analysis of stellar motions and their uncertainties from Gaia for use in targeting METI signals, which might be useful for SETI as well.

\section{Target and Frequency Selection (\thetargets\ papers)}

The Cosmic Haystack is large, and so narrowing down the best targets and frequencies to check is essential.

This problem of finding someone also looking for you is a game introduced to game theory by Thomas Schelling, who actually used \citet{cocconi_searching_1959} as an example of his point. In \citet{Wright2020_Planck} I described and formalized this connection of target and frequency selection as a hunt for ``Schelling points'' and applied the idea to the suggestion of \citet{PlanckUnits} that his eponymous units would be truly universal, known even to alien species. The resulting lists of frequencies to check (across the electromagnetic spectrum) are arguably still too anthropocentric and arbitrary, but are nonetheless small enough in number to add to the lists of ``magic frequencies'' to search at.

Another Schelling point is when to look.  \citet{Gray2020a} argued that an efficient transmission strategy is to broadcast continuously from a fixed point and direction on a planetary surface. The result would be that signals would repeat every planetary rotation. Gray's application here is to the Wow!\ Signal (and indeed this motivated \citet{Harp2020a} above) but more generally is that similar ``one-off'' signals should be prioritized for re-observation.  

We also need to know where to look. \citet{Caballero2020} attempted to identify the most likely stars in the Wow!\ Signal region of the sky to host life.  \citet{Kaltenegger2020} used the Gaia DR2 catalog to extend the work of \citet{Heller16} to identify new Earth Transit Zone stars.

Arguments for searches for artifacts within the solar system are as old as modern SETI \citep{Bracewell60}, but the idea seems to have new resonance these days. \citet{Gertz2020a} made one such argument, and includes some suggestions for how to search.

Developing ideas by \citet{Wolfe1985}, \citet{Turnbull03b}, and others, \citet{Gertz2020b} made a good argument that white dwarfs, evolved stars, and similar objects, while presumably poor targets for searches for biosignatures, are actually good targets for technosignature searches because they have given rise to a technological species that has outlived its host planet.

\citet{Cirkovic2020a} argued that M dwarfs are poor targets for radio SETI, because their frequent flaring would interfere with the development of radio technology on their planets (although I wonder whether the presence of a powerful and important radio source like a flaring star nearby would not \textit{increase} the likelihood of the development of such technology).

Finally, \citet{Lacki2020b} produced a remarkable catalog of ``one of everything,'' a list of an example of every major kind of astronomical object and superlative. Interesting in its own right beyond SETI, this ``exotica catalog'' for Breakthrough Listen also provides a target list for investigation into whether any of these presumably natural kinds of objects is actually an example of or influenced by technology on large scales.

\section{Development of Technosignatures (\thedevelopment\ papers)}

\subsection{The Axes of Merit}

Many technosignatures across a range of scales and detection strategies have been proposed, and it remains unclear which are the most promising.

The literature has long seen papers arguing for various criteria of merit when weighing technosignatures. Indeed, within two years of \citet{cocconi_searching_1959}'s foundational article arguing for radio communication as an ideal technosignature, \citet{dyson60}, \citet{Bracewell60}, and \citet{Schwartz61} all argued for others on various grounds. For a good example, see the discussion of \citet{Shostak2020}, who made the case for artifacts over radio transmissions.

At the NASA Technosignatures Workshop in 2018, Sofia Sheikh compiled many of these criteria into a single diagram she called the ``axes of merit'' for technosignature search.  This helpfully compiled various arguments about detectability, contrivance, longevity, and so on into a single, qualitative, somewhat subjective framework. \citet{Sheikh2020b} formalized this, and provided a software tool for creating one's own versions of the diagram for particular technosignatures.

\subsection{Dyson spheres, von Neumann probes, and spacecraft}

In \citet{Wright2020_Dyson} I reviewed the literature on Dyson Spheres, discussing the origins of the idea, their stability, and their thermodynamics. I also extended the AGENT framework developed in \citet{GHAT2} to calculate their observational signatures in the optical and infrared.  

\citet{Haliki2020} presented some calculations of how a Dyson sphere of self-replicating von Neumann probes might behave. \citet{Osmanov2020a} and \citet{Osmanov2020b} explored some various spectral and energetic properties of interstellar von Neumann machines.

Interstellar spacecraft might produce detectable signatures. \citet{Hoang2020} explored the detectability of mildly relativistic objects in the Solar System, concluding \textit{JWST} would be able to detect them.  \citet{Lingam2020b} explored how extremely luminous events such as supernovae or active galactic nuclei could power spacecraft with solar sails, and the observational consequences of such propulsion.

\subsection{Novel Technosignatures}

2020 brought us several new technosignatures to consider searching for.

In \citet{Hippke2020c} and \citet{Hippke2020b}, Michael Hippke began a series of papers developing an idea for how an interstellar communication network would likely be set up using stars as gravitational lenses. Such a network could have specific observational consequences which he will presumably flesh out in future installments of the series.

\citet{Jackson2020} explored several ideas for technosignatures including from spacecraft and megastructures, and including neutrinos and gravitational waves.  \citet{Abramowicz2020} also explored gravitational waves as a technosignature, in particular how the Milky Way's supermassive black hole could produce detectable gravitational waves if used as a power source. \citet{Santos2020} connected neutrinos detected during binary neutron star merger events to controversial claims that radioactive decay rates are influenced by neutrino flux, to suggest searching for decay rate fluctuations during such merger events as a deliberate signal.

\citet{Lingam2020a} suggested looking for hypervelocity stars as a technosignature (being the result of stellar engines). Finally, \citet{Lacki2020a} explored how and whether X-ray binaries could be used to power ``passive'' communication beacons with high detectability. 

\section{Theory of ETIs (\thetheory\ papers)}

Theory on the nature, scale, lifetime, and ubiquity of ETIs was one of the largest categories in 2020, something I suspect is true most years.

\subsection{The Drake Equation and the Fermi Paradox}

The most ubiquitous topics for theory papers in SETI are the Drake Equation and the Fermi Paradox. Indeed, the topic predates modern SETI, with the Fermi Paradox ostensibly going back to Fermi's lunchtime discussion in 1950 \citep{FermiParadox} \citep[and in truth much farther, see, e.g., Chapter 8 of][]{LingamBook} and early Drake-like calculations going back at least as far as \citet{Maunder1913} \citep[as][pointed out in a recent research note]{Lorenz2020}. I categorize papers here rather broadly, including papers on the ``Great Filter'' \citep{Filter} and similar considerations.

\citet{DeVisscher2020} argued from the Drake Equation that artificial intelligences should outnumber biological ones in the galaxy. \citet{Tsumura2020} used the history of extinctions on Earth to attempt to constrain the $f_i$ term in the Drake Equation. 
\citet{Kipping2020a} examined what life's early and intelligence's late emergence on Earth tells us about $f_l$ and $f_i$. \citet{Westby2020} performed a Drake Equation calculation with loose and strong priors based on Earth life (which they call the ``weak'' and ``strong'' Copernican limits). \citet{Zhang2020_Drake} provided a derivation of a form of the Drake Equation, and discussed it from the perspective of a transmitting species, in the context of the Arecibo message and fast radio bursts.

\citet{Prantzos2020} and \citet{Lares2020a} both performed Monte Carlo simulations of the rise and fall of technological life in the galaxy, exploring the likelihood of contact via communication given various assumptions. \citeauthor{Prantzos2020} emphasized the importance of the $L$ term in the Drake Equation, and \citeauthor{Lares2020a} emphasized that contact is most likely to occur just after the discovery of the relevant communicative technology.

\citet{Spada2020} extended the work of \citet{DeVito2019} arguing that our lack of contact with alien species is best explained if the growth rate of communicative species is logarithmic, or at best sub-linear.

\subsection{Categorization and Evolution of ETIs}

\citet{Kipping2020b} rigorously justified our expectation that when contact is made it will be with a species that has been technological for far longer than humans, and why this expectation is robust even if most technological species do not last very long. 

\citet{Dobler2020} discussed the assumptions he claims underlie ``orthodox'' radio SETI, in particular the problematic nature of assuming a common path of technological development among many species, and explores the importance of ``technological synchronicity.'' \citet{Panov2020} examined what post-Singularity ETIs might be like, with an eye towards their detectability. \citet{Gale2020} argued that the Singularity is both nigh and universal, and that SETI would therefore do best to search for post-biological intelligences exploiting features such as quantum entanglement for communication.\footnote{Their title, ``Will recent advances in AI result in a paradigm shift in Astrobiology and SETI?'' may be a rare violation of Betteridge's Law, which states that articles with questions for titles will argue the answer is ``no.''} 

\citet{Gray2020b} described the ``extended'' Kardashev scale, with humanity's energy supply at 0.72 on the scale, and discusses how it can be generalized beyond energy use to describe, for instance, information or population. \citet{Ivanov2020} described additional axes for Kardashev's scale to describe ETIs, including their degree of interaction and integration with their environment, with implications for their detectability.

\subsection{The Copernican and anthropic principles}

Closely related to the Drake Equation and Fermi Paradox are calculations concerning what, if anything, we can conclude from the facts of our own existence. The Copernican Principle argues, roughly, that we should not expect to discover that life on Earth is special in too many or extreme ways; and the anthropic principle cautions that our own status as observers is the result of a selection bias, limiting the utility of ourselves as a data point in statistical considerations.

 \citet{Cirkovic2020b} argued that typicality in time is not a well defined concept, and warned that we should use caution when deploying ``temporal Copernicanism.'' 

Following a study by \citet{Dayal2015} arguing that most habitable planets exist in elliptical galaxies, \citet{Whitmire2020} argued that our existence in a spiral galaxy therefore violates the Principle of Mediocrity. To resolve this conflict, he hypothesized reasons that elliptical galaxies may not be especially habitable after all.  

\citet{Olson2020} examined how the philosophical self-indication assumption and self-sampling assumption (which are forms of anthropic reasoning) affect our expectations for finding large extragalactic technologies.

Finally, \citet{Slijepcevic2020} argued that cognition is a biological universal, and that many anthropic arguments by astrophysicists are incorrectly and tacitly predicated on assumptions of human-like intelligence.

\section{Social aspects of SETI (\thesocial\ papers)}

The organization of the worldwide SETI community is a matter of ongoing discussion, and, as always, METI occupied a large part of it in 2020. 

\citet{Kellermann2020} published a history of NRAO that includes a nice discussion of the early days of SETI. \citet{Abrevaya2020} discussed SETI in the context of the history of astrobiology research in  Argentina, and \citet{Lingam2020c} discussed it in the context of the history of astrobiology.

\citet{Haqq-Misra2020b} had the clever idea to solve some of SETI's funding problems via a lottery bond.  \citet{Pasachoff2020} discussed the place of the Voyager Golden Records in intellectual history.  

\citet{Cortellesi2020} framed METI efforts within a broader ``continuum of astrobiological signaling'' and in the context of the ``SETI Paradox'' \citep[in which searches for communications will fail if all species only listen and never transmit, see][]{Zaitsev2006_Paradox}. \citet{Smith2020a} discussed the ethical considerations of METI in terms of informed consent.

\citet{Vidaurri2020} discussed METI and SETI, briefly, in the broader context of an anti-imperial approach to astrobiology and space exploration. \citet{Smith2020b} argued for a position of extreme passivity for any First Contact scenario, a position that echoes the ``Radical First Contact'' of \citet{omanreagan2018} in astrobiology.  

\citet{Wisian2020} argued that SETI practitioners need to be concerned about the geopolitical implications of success, presenting a \textit{realpolitik} analysis of the potentially deadly competition among states that would follow to secure a monopoly on the advanced scientific ideas that would flow from contact. (Their scenario is rather implausible, however, and a rebuttal is in the works).

\section{Looking Ahead}

\begin{displayquote}

\textit{I've got a feeling '21\\
Is going to be a good year}

---The Who, \textit{Tommy}

\end{displayquote}

2021 promises revival in many ways. We can look forward to the US Decadal Survey of Astronomy and Astrophysics, which we hope will include an explicit recommendation for funding for technosignature searches. Many students will finish courses in SETI at Penn State and UCLA, and we expect some will graduate with theses devoted to the topic. As vaccine availability becomes more widespread worldwide, we hope SETI conferences may begin again in-person, perhaps starting with the meeting of the IAA Permanent Committee on SETI at the International Astronautical Congress in Dubai in October. And we can hope that more new SETI initiatives are announced through traditional and nontraditional funding sources. 

I hope to have another review with an update on all of these items in a year or so, and of course we can all follow the field as it develops in the literature at the \href{https://seti.news}{seti.news} website.

\begin{acknowledgements}

I thank Manasvi Lingam for helpful comments and for providing several references I had missed. I thank Virginia Trimble for her her APxx series, and for her suggestion that I owe her a bottle of vermouth because of this paper, which I choose to interpret as an implicit endorsement (n.b.\ the debt has been paid). I thank the anonymous referees for their reviews. 

This research has made use of NASA's Astrophysics Data System Bibliographic Services. This research was supported by the Center for Exoplanets and Habitable Worlds and the Penn State Extraterrestrial Intelligence Center, which are supported by the Pennsylvania State University and the Eberly College of Science.




\end{acknowledgements}


\end{document}